\documentstyle[12pt,titlepage]{article}
\newlength{\extraspace}
\newlength{\extraspaces}
\catcode`\@=11

\def\numberbysection{\@addtoreset{equation}{section}
\def\theequation{\arabic{section}.\arabic{equation}}}

\newcommand{\be}{\begin{equation}
\addtolength{\abovedisplayskip}{\extraspaces}
\addtolength{\belowdisplayskip}{\extraspaces}
\addtolength{\abovedisplayshortskip}{\extraspace}
\addtolength{\belowdisplayshortskip}{\extraspace}}
\newcommand{\ee}{\end{equation}}
\newcommand{\ba}{\begin{eqnarray}
\addtolength{\abovedisplayskip}{\extraspaces}
\addtolength{\belowdisplayskip}{\extraspaces}
\addtolength{\abovedisplayshortskip}{\extraspace}
\addtolength{\belowdisplayshortskip}{\extraspace}}
\newcommand{\ea}{\end{eqnarray}}
\newcommand{\nonu}{\nonumber \\[.5mm]}

\newcommand{\R}{\bf R}

\newcommand{\tg}{\tilde{g}}

\newcommand{\tbeta}{\tilde{\beta}}

\begin{document}
\begin{titlepage}
\addtolength{\baselineskip}{.7mm}
\thispagestyle{empty}
\begin{flushright}
TIT/HEP--215 \\
NUP-A-93-4 \\
hepth@xxx/9303090\\
March, 1993
\end{flushright}
\vspace{2mm}
\begin{center}
{\large{\bf
Exact Beta Functions in the Vector Model and
Renormalization Group Approach }} \\[8mm]
%
{\sc Saburo Higuchi}
\footnote{{\tt e-mail: hig@phys.titech.ac.jp}, JSPS fellow}\\[3mm]
{\it Department of Physics, Tokyo Institute of Technology, \\[2mm]
Oh-okayama, Meguro, Tokyo 152, Japan} \\[4mm]
{\sc Chigak Itoi}
\footnote{\tt e-mail: itoi@phys.titech.ac.jp} \\[3mm]
{\it Department of Physics and Atomic Energy Research Institute, \\[2mm]
College of Science and Technology, Nihon University, \\[2mm]
Kanda Surugadai, Chiyoda, Tokyo 101, Japan} \\[4mm]
and \\[4mm]
{\sc Norisuke Sakai}
\footnote{\tt e-mail: nsakai@phys.titech.ac.jp} \\[3mm]
{\it Department of Physics,
Tokyo Institute of Technology, \\[2mm]
Oh-okayama, Meguro, Tokyo 152, Japan} \\[8mm]
{\bf Abstract}\\[5mm]
{\parbox{13cm}{\hspace{5mm}
The validity of the renormalization group approach for large $N$ is
clarified by using the vector model as an example.
An exact difference equation is obtained which relates
free energies for neighboring values of $N$.
The reparametrization freedom in field space provides infinitely many
identities which
reduce the infinite dimensional coupling
constant space to
that of finite dimensions.
The effective beta functions give exact values for the fixed points
and the susceptibility exponents.
}}
\end{center}
\vfill
\end{titlepage}
\setcounter{section}{0}
\setcounter{equation}{0}
%
%
The two-dimensional quantum gravity is important to study string
theories and is useful as a toy model for higher dimensional quantum
gravity.
The matrix model gives a discretized version
of the two-dimensional quantum gravity.
Exact solutions of the matrix model \cite{BIPZ}--\cite{GRMI}
have been obtained for two-dimensional quantum gravity coupled to
conformal matter with central charge $c \le 1$.
The result can also be understood by means of the continuum approach
\cite{DDK}--\cite{BEKL}.
Although several exact solutions of the matrix model have been
obtained, it is worth studying approximation schemes which
enable us to calculate critical coupling constants and
critical exponents for unsolved matrix models, especially for $c>1$.
In order to make use of such a scheme, we need to make sure
that the approximation method gives correct results for the
exactly solved cases.
\par
Recently Br\'ezin and Zinn-Justin have proposed a renormalization group
approach to the matrix model\cite{BRZJ}.
They drew an analogy between
the fixed point of the renormalization group flow
and the double scaling limit
of the matrix model \cite{GRMI}.
They observed that a change  $N \rightarrow N+\delta N$ can
be compensated by a change of coupling
constants $g \rightarrow g+\delta g$  in order
to give the same continuum physics.
They needed to enlarge the coupling constant space
as in the Wilson's renormalization group approach \cite{WIKO}.
Consequences of their approach have been examined by several groups
\cite{ALDA}.
A similar approach has been advocated previously for
the $1/N$ expansion in a somewhat different context \cite{CARL}.
In the case of the one-matrix model with $c\le 1$,
Br\'{e}zin and Zinn-Justin obtained reasonable results
 for the fixed point and the susceptibility exponents
in the first nontrivial approximation.
In order to demonstrate the validity of the renormalization
group approach, however, one should show that
the systematic improvement
of their approximate evaluation converges to the correct result.
To this end, it is important to study a model in which a
renormalization group equation can be derived exactly, even if
it is simpler than the matrix model.
The vector model has been proposed for a discretized
one-dimensional quantum gravity, in the same way as
the matrix model for a discretized
two-dimensional quantum gravity\cite{NIYO}--\cite{ZJ}.
\par
The purpose of our paper is to clarify the validity and the meaning
of the renormalization group approach by considering
the vector model.
For the vector model, we obtain an exact difference equation which
relates the free energy $-\log Z_{N-2}(g)$
to the free energy $-\log Z_N(g - 2\delta g)$ with
slightly different values of coupling constants.
We find that these coupling constant shifts $\delta g_k$ are
of order $1/N$ and occur in infinitely many coupling constants.
We also obtain infinitely many identities which
express the freedom to reparametrize the field space.
By using these identities, we can rewrite the flow
in the infinite dimensional coupling constant space
as an effective flow in the space of finite
number of coupling constants.
The resulting effective beta function determines the fixed
points and the susceptibility exponents.
The inhomogeneous
term in the effective renormalization group equation
serves to fix non-universal (analytic)
terms of the free energy.
To illustrate the procedure by an explicit example,
we analyze in detail the cases of one and two coupling constants which
give the first two
multicritical points $m=2$ and $3$.
We obtain the fixed points and the
susceptibility exponents which are in complete agreement with the exact
results.
\par
%
%
We first recall the renormalization group approach for the matrix model.
The partition function $Z_N(g)$ of the matrix model with a single
coupling constant $g$ for the quartic interaction is defined by an
integral over an $N\times N$ hermitian matrix $\Phi$
\be
Z_N(g)= \int d^{N^2} \Phi \exp \left[-N\left({1 \over 2}\Phi^2
+{g \over 4}\Phi^4 \right) \right].
\ee
The matrix model gives the random ``triangulation''
of two-dimensional
surfaces \cite{BIPZ}.
The $1/N$ expansion of the free energy $F(N, g)$
\be
 F (N, g)= -{1 \over N^2}\log Z_N(g)
= \sum_{h=0}^{\infty} N^{-2h} f_h(g)
\ee
distinguishes the contributions $f_h$ from the surface with $h$
handles.
In the double scaling limit
\be
 N \rightarrow \infty,   \qquad g \rightarrow g_*,  \qquad
{\rm with}   \quad   N^{2 \over \gamma_1}(g - g_*)
\quad {\rm fixed,}
\label{eqn:doublescalinglimit}
\ee
the singular part of the free energy satisfies the scaling law
 \cite{DDK}
with the susceptibility exponent
$\gamma_0 + h \gamma_1$
linear in the number of handles
$h$
\ba
 f_h (g) &\!\!\! = &\!\!\!
(g - g_*)^{2 - \gamma_0-\gamma_1 h} a_h + \cdots, \nonu
 F (N,g) &\!\!\! = &\!\!\! (g - g_*)^{2 - \gamma_0}
f \bigl( N^{2 \over \gamma_1}(g - g_*) \bigr).
\ea
The functional form of $f$ can be determined by a nonlinear
differential equation (string equation) \cite{GRMI}.
The continuum limit is achieved at the critical point
$g \rightarrow g_*$ where the average number of triangles
diverges.
The double scaling limit suggests that we may draw an analogy
between $N^{2 \over \gamma_1}$ and the momentum cut-off $\Lambda^2$.

It has been proposed that the free energy $F(N,g)$ of the matrix model
satisfies the following renormalization group equation
 \cite{BRZJ}
\be
\biggl[N{\partial \over \partial N}
-\beta (g) {\partial \over \partial g}
+\gamma(g)\biggr] F (N, g)=r(g),
\ee
where $\beta(g)$ is called the beta function.
The anomalous dimension and the inhomogeneous term are
denoted as $\gamma(g)$ and $r(g)$ respectively.
A fixed point $g_*$ is given by a zero of the beta function.
The exponents of the
double scaling limit can be given by the derivative of the
beta function at the fixed point
$\gamma_1 = 2/ \beta'(g_*), \; \gamma_0 = 2- \gamma(g_*)/\beta'(g_*)$.
As a practical method to obtain
the beta function,
Br\'ezin and Zinn-Justin proposed an approximate evaluation of the
beta function by integrating over a part of degrees of freedom of the
$N\times N$ matrix
 \cite{BRZJ}, \cite{CARL}.
They found a reasonable result at the first nontrivial order.
We have computed higher orders of the matrix model
following their approximation methods,
but we find no signs of improvement as we proceed to higher orders.
Before discussing the result of the matrix model, we find it
more illuminating to study the case of the vector model where we
can clarify the situation more fully.

The partition function of the $O(N)$ symmetric vector model is given by
\begin{equation}
  Z_N(g) = \int d^N \phi
  \exp \left[ -N \sum_{k=1}^{\infty} \frac{g_k}{2k} (\phi^2)^k\right],
  \label{eqn:vector_action}
\end{equation}
where $\phi$ is an $N$ dimensional real vector
\cite{NIYO} -- \cite{ZJ}.
Here we introduce infinitely many coupling constants $g_k$,
since we need all possible induced interactions after a
renormalization group transformation even if we start with a few
coupling constants only.
The $1/N$ expansion of the logarithm of the partition function
gives contributions from $h$ loops as terms with $N^{1-h}$.
The vector model  has the double scaling limit
$N \rightarrow \infty $ with $N^{1/\gamma_1}(g-g_\ast)$ fixed,
where
the singular part of the free energy satisfies the scaling law
\cite{NIYO} -- \cite{ZJ}
\be
 -\log \left.\biggl[{Z_N(g) \over Z_N(g_1=1,g_k=0 \; (k\geq2))}
 \biggr]\right|_{sing}
= \sum_{h=0}^{\infty} N^{1-h}
(g - g_*)^{2 - \gamma_0-\gamma_1 h} a_h + \cdots.
\ee
One should note that the power of $1/N$ and the definition of the
susceptibility exponents are slightly different from the
matrix model.
Therefore the relation between these susceptibility exponents and the
derivative of the beta function becomes
\be
\gamma_1 = {1 \over \beta'(g_*)}, \qquad
\gamma_0 = 2- {\gamma(g_*) \over \beta'(g_*)}.
 \label{eqn:vectorgamma_and_beta}
\ee

In the spirit of the
approximation method of ref.\cite{BRZJ}, we can integrate over
the $N+1$-th component $\alpha$ of the  vector $\phi_{N+1}$
\ba
\phi_{N+1} &\!\!\! = &\!\!\! (\phi_N, \alpha), \nonu
  Z_{N+1}(g)
    &\!\!\! = &\!\!\! \int d^N \phi_N d\alpha
            \exp \left[ -(N+1) \sum_{k=1}^{\infty}
                 \frac{g_k}{2k} (\phi_N^2+\alpha^2)^k\right].
\ea
Neglecting higher order terms in $1/N$, we obtain
\be
\frac{Z_{N+1}(g) }{Z_{N+1}(g_1=1,g_k=0\;(k\geq2))}
= \int d^N \phi_N
  \exp \left[ -N \sum_{k=1}^{\infty} \frac{g_k+\delta g_k}{2k}
       (\phi_N^2)^k + O\left({1 \over N}\right) \right],
   \label{eqn:approxbeta}
\ee
where the shifts $\delta g_k$ of the coupling constants are found to be
\begin{equation}
  \sum_{k=1}^{\infty} \frac{g_k}{k} x^k
  + \log \left(\sum_{k=1}^{\infty} \frac{g_k}{g_1} x^{k-1}\right)
  =
  N \sum_{k=1}^{\infty} \frac{\delta g_k}{k}    x^k.
   \label{eqn:shift}
\end{equation}

Now we shall show that the shift
$\delta g_k$ can be evaluated exactly in the vector model
without using approximation methods such as described above.
To demonstrate that the result is exact, we start with the partition
function $Z_{N-2}(g)$ .
After integrating over angular coordinates in ${\R}^{N-2}$,
we perform a partial integration in the radial coordinate
$x=\phi^2$
\begin{eqnarray}
  Z_{N-2}(g)
    &\!\!\! = &\!\!\! \frac{\pi^{\frac{N}{2}-1}}{\Gamma(\frac{N}{2}-1)}
          \int_{0}^{\infty} dx\;
       x^{N/2-2} \exp\left[-(N-2)\sum_{k=1}^{\infty}
          \frac{g_k}{2k}x^k\right] \nonu
    &\!\!\! = &\!\!\! \frac{\pi^{\frac{N}{2}-1}}{\Gamma(\frac{N}{2}-1)}
          \int_{0}^{\infty} dx\;
       x^{N/2-1} \left(\sum_{k=1}^{\infty} g_k x^{k-1}\right)
         \exp \left[-(N-2)\sum_{k=1}^{\infty} \frac{g_k}{2k}x^k\right].
   \label{eqn:radial_int}
\end{eqnarray}
Identifying the  right hand side with
$((N-2) g_1/ 2\pi) Z_{N}(g - 2 \delta g)$,
we obtain an exact difference equation for the
logarithm of the partition function
\begin{eqnarray}
\lefteqn{[(-\log Z_{N}(g))-(-\log Z_{N-2}(g))]
- \log\frac{(N-2) g_1}{2\pi}}
 \nonumber \\
 &\!\!\! =&\!\!\! - [ ( -\log Z_{N} ( g - 2 \delta g))
 - ( -\log Z_{N}(g))].
  \label{eqn:difference_equation}
\end{eqnarray}
We find that the shifts $\delta g_k$ of the coupling constants are
exactly identical to the result (\ref{eqn:shift})
of the approximate evaluation.
We would like to stress
that no approximation is employed to obtain
eq. (\ref{eqn:difference_equation}).
Therefore we can infer that
the vector model offers an example to justify the approximate evaluation
method of ref.\cite{BRZJ}.
\par
In the $N \rightarrow \infty$ limit, we can obtain a
differential equation from the exact
difference equation (\ref{eqn:difference_equation})
\begin{equation}
  \frac{\partial}{\partial N}( -\log Z_{N} ( g ))
        - \frac{1}{2} \log \frac{Ng_1}{2\pi}
  =
  \sum_{k=1}^{\infty}  \delta g_k \frac{\partial}{\partial  g_k}
( -\log Z_{N} ( g )).
  \label{eqn:differential_equation}
\end{equation}
One can bring the quadratic
term in the potential to the standard form $\phi^2/2$
since $g_1$ can be absorbed
by a rescaling $g_1 \phi^2 \rightarrow \phi^2$. We have
\begin{equation}
 Z_N(g_1,g_2,g_3,\ldots) = g_1^{-N/2} Z_N(1, g_2/g_1^2,g_3/g_1^3,\ldots).
\label{rescalingid}
\end{equation}
Therefore it is convenient to use the
rescaled coupling constants $\tilde g_k$ together with $g_1$
as independent coupling constants
\begin{equation}
 \tg_k = g_k / g_1^k.
\end{equation}
We shall define the free energy for the vector model
\begin{equation}
F(N,\tg) = - \frac{1}{N} \log Z_N(g)
- \frac{1}{2} \log \frac{Ng_1}{2\pi}.
\end{equation}
If we use $g_1, \tg_2, \tg_3, \cdots$  as independent coupling
constants, we find from the rescaling identity (\ref{rescalingid})
that the partition function $Z_N$ depends on $g_1$ only through the
factor $g_1^{-N/2}$.
Therefore, the free energy $F(N, \tilde g)$ is independent of
$g_1$ and is a function of $\tg_k \; (k \geq 2)$ only.
\par
We denote the partial derivatives with respect to
$g_1, \tilde g_2, \tilde g_3, \cdots$ by $|_{\tilde g}$, and
those with $g_1, g_2, g_3, \cdots$
by $|_{g} $
\begin{equation}
 \left.\frac{\partial}{\partial g_1}\right|_g
=\left.\frac{\partial}{\partial g_1}\right|_{\tg}
   - \sum_{k=2}^{\infty} k \frac{\tg_k}{g_1}
               \left.   \frac{\partial}{\partial    \tg_k}\right|_{\tg},
\qquad
 \left.\frac{\partial}{\partial g_k}\right|_g
= \frac{1}{g_1^k} \left. \frac{\partial}{\partial \tg_k}\right|_{\tg} .
\end{equation}
Thus we obtain a renormalization group equation for the free
 energy $F$
\begin{equation}
\left[N \frac{\partial}{\partial N}
  -
  \sum_{k=2}^{\infty}
N \left( \frac{\delta g_k}{g_1^k} - \frac{\delta g_1}{g_1} k \tg_k\right)
\left.     \frac{\partial}{\partial \tg_k} \right|_{\tg}
 +1   \right]F(N,\tg)
= N \delta g_1 \frac{1}{2g_1}
      - \frac{1}{2}   .
 \label{eqn:renorm_all_coupling}
\end{equation}
Eq.(\ref{eqn:renorm_all_coupling}) shows that the anomalous
dimension is  given by
\be
  \gamma(\tg) =1,
\label{eqn:anomalousdim}
\ee
which implies a relation between two susceptibility exponents
\be
  \gamma_0 + \gamma_1 = 2.
\ee
We read off the beta functions in the rescaled coupling
constants as
\begin{equation}
\tbeta_k(\tg)
=N \left(
     \frac{\delta g_k}{g_1^k} - \frac{\delta g_1}{g_1} k
     \tilde{g_k}
     \right).
   \label{eqn:rescaled_beta}
\end{equation}
These beta functions can be evaluated
explicitly using eq.(\ref{eqn:shift}) as
\begin{eqnarray}
  \tbeta_2(\tg) &\!\!\! = &\!\!\! -\tg_2 - 3 \tg_2^2 + 2 \tg_3, \nonu
  \tbeta_3(\tg) &\!\!\! = &\!\!\! -2\tg_3
  + \tg_2^3 -6\tg_2 \tg_3 + 3 \tg_4, \nonu
  \tbeta_4(\tg) &\!\!\! = &\!\!\!-3\tg_4 - \tg_2^4 + 4\tg_2^2\tg_3 -
  2\tg_3^2 - 8 \tg_2 \tg_4 + 4 \tg_5,  \nonu
                &\!\!\! \vdots &\!\!\! .
  \label{eqn:explicit_beta}
\end{eqnarray}

Let us investigate the simultaneous zero of the beta functions
$\tilde \beta_k=0 \; (k\geq 2)$ in the spirit of ref.\cite{BRZJ}.
The first condition $\tilde \beta_2 = 0$ gives $\tilde g_3$ in terms of
$\tilde g_2$.
The second condition $\tilde \beta_3=0$ determines $\tilde g_4$
in terms of $\tilde g_2$ and $\tilde g_3$, and hence
in terms of $\tilde g_2$.
As can be seen in eq.(\ref{eqn:explicit_beta}),
$\tbeta_k(\tilde g)$ turns out to be a sum of $k \tilde g_{k+1}$
and a polynomial in $\tilde g_j, j\le k$.
Therefore we find that there always exists a solution of
$\tilde \beta_k=0 \; (k\geq 2)$
for each given value of the coupling constant $\tg_2$.
In the following, we shall show that this strange result
of the apparent existence of
the one-parameter family of fixed points is due to a misinterpretation
of the renormalization group flow.
\par
The key observation is the ambiguity to identify the renormalization
group flow in the coupling constant space.
Though the above equation (\ref{eqn:renorm_all_coupling}) seems to
describe a renormalization group flow in the infinite dimensional
coupling constant space,
the direction of the flow is in fact ambiguous
because all the differential operators $(\partial/\partial\tilde{g_k})$
are not linearly independent as we will see shortly.
To see this, note that the partition function (\ref{eqn:vector_action})
is invariant under reparametrizations of the integration variable
$\phi$.
Since the model is $O(N)$ invariant, we can obtain new informations
only from reparametrizations of the radial coordinate $x=\phi^2$.
Since the radial coordinate should take values on the half real line,
we consider the most general reparametrization which keeps the
integration range $[0,\infty)$
\begin{equation}
 x = y \left( 1+ \sum_{j=0}^{\infty} \varepsilon_j y^j\right),
 \label{eqn:reparametrization}
\end{equation}
where $\varepsilon_j$'s are infinitesimal parameters.
Substituting (\ref{eqn:reparametrization})
in (\ref{eqn:radial_int}) and differentiating with respect to
$\varepsilon_j$'s,
we obtain a family of identities
\begin{eqnarray}
 &\!\!\! &\!\!\! L_j Z_N(g) = 0  \mbox{\ \ \ } \qquad ( j \geq 0), \\
 &\!\!\! &\!\!\! L_j = \sum_{\ell=j+1}^{\infty}g_{\ell-j}\ell
           \left.\frac{\partial}{\partial g_\ell}\right|_g
        - \left( 1+ \frac{2j}{N}\right)
          \left. j \frac{\partial}{\partial g_j}\right|_g
           + \frac{N}{2} \delta_{j,0}.
\end{eqnarray}
The differential operators $L_j$ constitute  half of the
Virasoro algebra.
We can show that this algebra is identical with the one found in
\cite{NIYO} and \cite{DVKO}.

It is more useful to use the rescaled coupling constants $\tilde g_k$.
The reparametrization identity corresponding to $\varepsilon_0$ is
nothing but the infinitesimal form of the rescaling identity
(\ref{rescalingid}) and reads
\be
\left.\frac{\partial}{\partial g_1}\right|_{\tg} F(N,\tg)  =  0.
\ee
In terms of the rescaled coupling constants,
the reparametrization identity corresponding to $\varepsilon_1$ is
given as
\be
-\frac{N+2}{2N}+
 \sum_{l=2}^{\infty}
\left\{\left(1+\frac{2}{N}\right)\tg_l+\tg_{\ell-1}\right\}
\left. \ell \frac{\partial}{\partial \tg_\ell}\right|_{\tg} F(N,\tg) = 0,
\label{eqn:firstreparametid}
\ee
The reparametrization identities corresponding to $\varepsilon_j$ is
given by
\be
\left.\left\{-\left(1+\frac{2j}{N}\right)
j \frac{\partial}{\partial \tg_j}
 + \sum_{l=j+1}^{\infty} \tg_{\ell-j}
  \ell \frac{\partial}{\partial \tg_\ell}
\right\}\right|_{\tg} F(N,\tg) = 0.
\label{eqn:jreparametid}
\ee
We see that derivatives of the free energy in terms of
infinitely many coupling constants $\tilde g_k$
are related by
infinitely many reparametrization identities.
Thus one can expect that only a finite number of derivatives are
linearly independent.
The exact difference equation (\ref{eqn:difference_equation})
combined with the reparametrization identities
(\ref{eqn:firstreparametid}) and (\ref{eqn:jreparametid})
constitute the complete set
of equations to characterize the renormalization group flow in our
approach.

To illustrate the use of the reparametrization identities,
we shall first take the case of a single coupling constant.
Let us consider a point in the coupling constant space
\be
(g_1, \tg_2, \tg_3, \tg_4, \ldots) = (g_1,\tg_2,0,0, \ldots).
\label{eqn:m2subspace}
\ee
At this point, the $j$-th identity relates
$\partial F / \partial \tg_{j+2}$ to
$\partial F / \partial \tg_{j+1}$ except for the case of $j=1$
where an extra constant term is present.
Therefore we can
express $\partial F / \partial \tg_k \; \; (k \geq 3)$ in terms
of $\partial F / \partial \tg_2 $ by
solving these reparametrization identities recursively
in the one-dimensional subspace (\ref{eqn:m2subspace}) of coupling
constants.
It is most convenient to organize the solution in powers of $1/N$.
We find explicitly at leading order
\begin{equation}
\frac{\partial F}{\partial \tg_k}
= B_k \frac{\partial F}{\partial \tg_2}  + R_k + O(N^{-1}),
\end{equation}
where
\begin{eqnarray}
B_k &\!\!\! = &\!\!\! \frac{(1-\Delta)^2}{16\sqrt{\Delta}}
         \left\{
       \left(\frac{2}{1-\sqrt{\Delta}}\right)^k
     - \left(\frac{2}{1+\sqrt{\Delta}}\right)^k
        \right\}           \frac{2}{k},\\
R_k &\!\!\! = &\!\!\! - \frac{1}{\sqrt{\Delta}}
         \left\{
        \left(\frac{2}{1-\sqrt{\Delta}}\right)^{k-2}
     - \left(\frac{2}{1+\sqrt{\Delta}}\right)^{k-2}
         \right\}           \frac{1}{2k},\\
\Delta &\!\!\! = &\!\!\! 4 \tg_2 + 1.
\end{eqnarray}
These solutions at leading order are sufficient to
eliminate $\partial F/\partial \tg_k  \; \; ( k \geq 3) $ in favor of
$\partial F/\partial \tg_2$ in (\ref{eqn:renorm_all_coupling}).
Thus we obtain a renormalization group equation with
the effective beta function $\beta^{\mbox{eff}}(\tg_2)$
and the inhomogeneous term $r(\tilde g_2)$
\begin{equation}
\left[N\frac{\partial}{\partial N}
- \beta^{\mbox{eff}}(\tg_2) \frac{\partial}{\partial \tg_2} + 1 \right]
 F(N,\tg_2) = r( \tg_2),
  \label{eqn:effective_rge}
\end{equation}
where
\begin{eqnarray}
\beta^{\mbox{eff}}(\tg_2)
&\!\!\! = &\!\!\! \sum_{k=2}^{\infty} B_k
\tbeta_k(\tilde g_2, \tilde g_k=0,k\geq 3)
\nonu
&\!\!\! = &\!\!\! {1 \over 4}
\left[1-\Delta - {(1-\Delta)^2 \over 2\sqrt{\Delta}}
\log \left({1+\sqrt{\Delta} \over 1-\sqrt{\Delta}}\right)\right]
\nonu
&\!\!\! = &\!\!\! \frac{2}{3} \left( \tg_2 + \frac{1}{4}\right)
     - \frac{32}{15} \left(\tg_2 + \frac{1}{4}\right)^2
       + \cdots,
\label{eqn:effectivebeta}
\\
r(\tg_2)
&\!\!\! = &\!\!\!
\frac{N}{2} \delta g_1 - \frac{1}{2}
+ \sum_{k=2}^{\infty} R_k
\tbeta_k(\tilde g_2, \tilde g_k=0,k\geq 3)
\nonu
&\!\!\! = &\!\!\! {1 \over 2\sqrt{\Delta}}
\left[\left({1+\sqrt{\Delta} \over 2}\right)^2
\log \left({1+\sqrt{\Delta} \over 2}\right)
-\left({1-\sqrt{\Delta} \over 2}\right)^2
\log \left({1-\sqrt{\Delta} \over 2}\right)\right]
\nonu
&\!\!\! = &\!\!\! \frac{1}{4} + \frac{1}{2} \log \frac{1}{2}
       + \frac{1}{3}\left( \tg_2 + \frac{1}{4}\right)
       + \frac{2}{15}\left(\tg_2 + \frac{1}{4}\right)^2 + \cdots.
\label{eqn:inhomogeneousterm}
\end{eqnarray}

The effective beta function
$\beta^{\mbox{eff}}(\tilde g_2)$ exhibits a zero at
$\tg_2 = -1/4$.
Furthermore, we can calculate the susceptibility exponents from
the derivative of the
effective beta function
by using (\ref{eqn:vectorgamma_and_beta}) and (\ref{eqn:anomalousdim})
\be
\gamma_1 = {1 \over \beta'(\tg_{2*})} = \frac{3}{2}, \qquad
\gamma_0 = 2- {\gamma(\tg_{2*}) \over \beta'(\tg_{2*})}
= \frac{1}{2}.
\ee
The fixed point and the susceptibility exponent are in complete
agreement with the exact results for the $m=2$ critical point of
the vector model corresponding to pure gravity \cite{NIYO} -- \cite{AMP}.
The beta function $\beta^{\mbox{eff}}(\tilde g_2)$ has also a trivial
fixed point at $\tg_2 = 0$, which is ultraviolet unstable since
$\partial \beta^{\mbox{eff}}/\partial \tilde g_2 (\tilde g_2=0)=-1<0$.

We can extract the complete information from the renormalization
group flow in our approach, namely the exact difference equation
and the reparametrization identities
by a systematic expansion of the free energy in powers of $1/N$
\begin{equation}
F(N,\tg_2) = \sum_{h=0}^{\infty} N^{-h} f_h(\tg_2).
\end{equation}
The free energy at leading order
$f_0$  satisfies an ordinary differential equation exactly,
\begin{equation}
  f_0(\tg_2) -
\beta^{\mbox{eff}}(\tg_2)\frac{\partial f_0}{\partial \tg_2}(\tg_2)
 = r(\tg_2).
 \label{eqn:diff_leadingN}
\end{equation}
where the effective beta function and the inhomogeneous term are
those given in eqs.(\ref{eqn:effectivebeta}) and
(\ref{eqn:inhomogeneousterm}).
We see immediately that the general solution is given by a sum
of an arbitrary multiple of the solution of the homogeneous equation and
a particular solution of the inhomogeneous equation.
Since both the effective beta function $\beta^{\mbox{eff}}(\tg_2) $
and the inhomogeneous term $r(\tilde g_2)$ are analytic in $\tg_2$
around the fixed point $\tg_2=-1/4$,
the singular behaviour of $f_0$
comes from the solution of the homogeneous equation.
It is important to notice that the singular term
corresponding to the continuum physics (the so-called universal term)
is specified by the beta function alone.
Their normalization, however, cannot be obtained from the
renormalization group equation.
The analytic contributions are determined by the
inhomogeneous term and the effective beta function.
\par
To obtain the full information for the free energy up to the order
$N^{-h}$, we should
write down the solution of the recursive reparametrization
identities (\ref{eqn:firstreparametid}) and (\ref{eqn:jreparametid})
up to the order $N^{-h}$ explicitly.
By inserting the solution into
the exact difference equation (\ref{eqn:difference_equation})
and by expanding it up to the power $N^{-h}$, we
obtain an ordinary differential equation for $f_h$
\be
  (1-h)f_h(\tg_2) -
\beta^{\mbox{eff}}(\tg_2)\frac{\partial f_0}{\partial \tg_2}(\tg_2)
= r_h(\tg_2).
\ee
We see that the effective beta function is common to all $h$,
whereas the inhomogeneous terms $r_h$ depend on $h$.
Therefore we find the singular part of the free energy
is determined by the beta function up to a
normalization $a_h$
\ba
f_h(\tg_2) &\!\!\! = &\!\!\!
f_h(\tg_2)_{sing}+f_h(\tg_2)_{analytic} \nonu
f_h(\tg_2)_{sing} &\!\!\! = &\!\!\!
(\tg_2 - \tg_{2*})^{2 - \gamma_0-\gamma_1 h} a_h + \cdots.
\ea
If we sum over the contributions from various $h$, we find that
the renormalization group equation determines the combinations of
variables appropriate to define the double scaling limit.
On the other hand, the functional form of the scaled variable is
undetermined corresponding to the undetermined normalization factor
$a_h$ for the singular terms of each $h$
\ba
F(N, \tilde g) &\!\!\! = &\!\!\!
\sum_{h=0}^{\infty} N^{-h}f_h(\tg_2)_{sing}
=\sum_{h=0}^{\infty} N^{-h}
(\tg_2 - \tg_{2*})^{2 - \gamma_0-\gamma_1 h} a_h \nonu
&\!\!\! = &\!\!\! (\tg_2 - \tg_{2*})^{2- \gamma_0}
f\bigl(N^{1/\gamma_1}(\tg_2-\tg_{2*})\bigr)_{sing}.
\ea

It has been known in the exact solution of the vector model
that a nonlinear differential equation
determines the functional form
$f\left(N^{1/\gamma_1}(\tilde g_2-\tg_{2*})\right)_{sing} $
of the scaled variables.
The equation is called the string equation,
or the $L_{-1}$ Virasoro constraint \cite{GRMI}, \cite{FKN}.
The above consideration means that the complete set of
our renormalization group flow equations, namely the exact difference
equation and the reparametrization identities, does not give
the $L_{-1}$ constraint for the universal singular terms.
This is in accord with our objective of the renormalization
group approach: to obtain at least fixed points and critical
exponents even if the exact solution is not available.
On the other hand, our equations give informations
on the non-universal analytic terms.

We can extend the analysis to the more general situation of finitely
many coupling constants.
Let us take
\be
g_1 = 1, \qquad
\tilde g_k \not= 0 \quad ( 2 \le k \le m), \qquad
\tilde g_k = 0 \quad ( k \ge m+1).
\ee
In this subspace, each reparametrization identity involves
only finite number of derivatives.
Moreover, the $j$-th identity relates derivatives in terms of
coupling constants $\tilde g_{j+1}, \tilde g_{j+2}, \cdots,
\tilde g_{j+m}$.
Therefore we can solve this identity to rewrite
$\partial F / \partial \tg_{j+m}$ in terms of
$\partial F / \partial \tg_k \; \; (j+1 \leq k \leq j+m-1)$.
By successively using these identities,
we find that there are precisely the necessary
number of recursion relations to express
$\partial F / \partial \tg_k \; \; (k \geq m+1)$
in terms of $\partial F / \partial \tg_k \; \; (2 \leq k \leq m)$.
Consequently, we can reduce the renormalization group equation
effectively in the space of the finite number of coupling constants
$\tilde g_k \;  (2\le k \le m)$
\begin{equation}
\left[N\frac{\partial}{\partial N}
- \sum_{k=2}^{m}
 \beta_k^{\mbox{eff}}(\tg_2,\ldots,\tg_m)
\frac{\partial}{\partial \tg_k}
+ 1 \right]
 F(N,\tg_2,\ldots,\tg_m) = r( \tg_2,\ldots,\tg_m).
\end{equation}
The $m$-th multicritical point
should be obtained as a simultaneous zero
of all the beta functions
$\beta_2^{\mbox{eff}}=\cdots=\beta_m^{\mbox{eff}}=0$.
The susceptibility exponent is given by eigenvalues of the
matrix of derivatives of beta functions
$\Omega_{ij}=\partial \beta_i^{\mbox{eff}}/\partial \tg_j$
 at the fixed point,
which is an $(m-1) \times (m-1)$ real matrix.

To illustrate the multi-coupling case explicitly, we take
the case of two coupling constants $m=3$.
We can solve the reparametrization identities recursively to
leading order in $1/N$, and combine
$\tilde \beta_k(\tilde g_2, \tilde g_3, \tilde g_k\!=\!0 \; (k\ge 4))$
to obtain the effective beta functions
\begin{eqnarray}
\beta_2^{\mbox{eff}}(\tg_2,\tg_3)
&\!\!\! = &\!\!\! \frac{-2 \alpha\beta\gamma}
{(\alpha\beta + \beta\gamma +\gamma\alpha)^2}
\frac{(\beta^3 - \gamma^3)\log \alpha + \mbox{cyclic}}%
{(\alpha-\beta)(\beta-\gamma)(\gamma-\alpha)}+
\frac{\alpha + \beta + \gamma}{\alpha\beta + \beta\gamma
+\gamma\alpha}, \\
\beta_3^{\mbox{eff}}(\tg_2,\tg_3)
&\!\!\! = &\!\!\! \frac{3 \alpha\beta\gamma}
{(\alpha\beta + \beta\gamma +\gamma\alpha)^2}
\frac{(\beta^2 - \gamma^2)\log \alpha + \mbox{cyclic}}%
{(\alpha-\beta)(\beta-\gamma)(\gamma-\alpha)}
- \frac{2}{\alpha\beta + \beta\gamma +\gamma\alpha},
  \label{multibeta}
\end{eqnarray}
where $\alpha,\beta,\gamma$ are the three roots of the cubic equation
\begin{equation}
  \tg_3 x^3 + \tg_2 x^2 + x -1 = 0.
 \label{eqn:saddlepointeq}
\end{equation}

We find three fixed points as the simultaneous zeros of
$\beta_2^{\mbox{eff}}$ and $ \beta_3^{\mbox{eff}} $.
All of them turn out to be on the $m=2$ critical line
\begin{equation}
\tg_3 = - \frac{1}{27} [ 2+ 9\tg_2 \pm 2( 1+ 3 \tg_2)^{3/2}].
  \label{eqn:critical_line}
\end{equation}
The first fixed point is at
$(\tg_2, \tg_3)=(-1/3,1/27)$.
This result agrees with the exact value of the coupling
constants at the $m=3$ critical point.
We have evaluated the derivative matrix
$\Omega_{ij}=\partial \beta_i^{\mbox{eff}}/\partial \tg_j$
by expanding the effective beta function around the fixed point
\begin{equation}
\Omega_{ij}
\left(\tilde g_2=-{1 \over 3},\tilde g_3={1 \over 27}\right)=
\left(
  \begin{array}{cc}
    \frac{11}{10} & \frac{9}{5} \\
    -\frac{7}{60} & \frac{3}{20}\\
  \end{array}
  \right).
\end{equation}
One of the eigenvalues of the derivative matrix is 3/4 which gives
the exact susceptibility exponent $\gamma_1 = 4/3$,
while the other eigenvalue is 1/2 which gives an analytic term.
The corresponding eigenvectors also agree with the exact solution.
As we stressed before, we can regard $N^{1/\gamma_1}$ as an
ultraviolet cut-off $\Lambda^2$ in taking the double scaling
limit.
Since both eigenvalues are positive,
the $m=3$ fixed point is ultraviolet stable.

The second fixed point is at $\tg_2= -1/4$ and $\tg_3 = 0$.
We obtain the derivative matrix
$\Omega_{ij}=\partial \beta_i^{\mbox{eff}}/\partial \tg_j$
at $(\tg_2,\tg_3)=(-\frac{1}{4},0)$
\begin{equation}
  \Omega_{ij}
 \left(\tilde g_2=-{1 \over 4},\tilde g_3=0\right)
  = \left(
  \begin{array}{cc}
    \frac{2}{3} & \frac{7}{3} \\ 0 & - \frac{1}{2}\\
  \end{array}
\right), \end{equation}
which has eigenvalues $2/3$ and $-1/2$.
The eigenvalue $2/3$ correctly gives the $m=2$ susceptibility exponent
$\gamma_1=3/2$.
The fixed point is attractive in the direction
$(\delta \tg_2,  \delta \tg_3) \propto (1,0)$
and repulsive in the direction
$(\delta \tg_2,  \delta \tg_3) \propto (-2,1)$,
which are the eigenvectors for the eigenvalues $2/3$ and $-1/2$,
respectively.
The repulsive direction is tangential to the $m=2$ critical line
(\ref{eqn:critical_line}) at the fixed point
$(\tg_2, \tg_3) = (-1/4,0)$.
\par
The third fixed point is the trivial fixed point at the origin
$(\tg_2, \tg_3) = (0,0)$.
Actually we find that the effective beta function
$\beta_2^{\mbox{eff}}(\tg_2, \tg_3=0)$ along the $\tg_2$ axis is
identical to the effective beta function $\beta_2^{\mbox{eff}}(\tg_2)$
in eq.(\ref{eqn:effectivebeta}) obtained in the single coupling
constant case.
This is because $\beta_3^{\mbox{eff}}(\tg_2, \tg_3=0)=0 $ along the
$\tg_2$ axis.

The critical line (\ref{eqn:critical_line})
is a trajectory of the renormalization group flow. {}From the
trivial fixed point $(\tg_2,\tg_3) = (0,0)$,
the flow goes to the  $m=3$ fixed point $(\tg_2,\tg_3) = (-1/3,1/27)$
along the critical line as $N \rightarrow \infty$.
Starting from a point on the critical line at the left of the $m=2$
nontrivial fixed point $(\tg_2,\tg_3) = (-1/4,0)$, the flow also goes
to the $m=3$ fixed point.
On the other hand, the flow goes to infinity if it starts from a point
on the critical line at the right of the $m=2$ nontrivial fixed point.
We notice that all three fixed points are isolated zeros of the
effective beta functions, since the derivative matrix
is non-degenerate at these fixed points.
Explicit evaluation
shows that the effective beta function does not vanish at any other
point on the $m=2$ critical line.

%
As we described earlier, a naive application of the
renormalization group approach to matrix models has given us results
which show no improvement as we go to higher orders.
In view of our analysis of the exact
result on the flow of the coupling constants for
the vector model,
one should not be surprised by these results on matrix models.
We should look for an effective renormalization group flow by
taking account of the reparametrization freedom in the field space.
Although the above results are derived exactly only for the
vector model, we have a similar ambiguity to identify the
renormalization group flow in the coupling constant space for
matrix models.
For instance, there are identities which form a representation
of Virasoro algebra in matrix models \cite{FKN}.
There has been some work to interpret the Virasoro constraints as
reparametrization identities in the field space in the matrix
model\cite{ITOYAMAMATSUO}.
We are trying to devise a way to implement
our idea of the effective renormalization group flow through the
application of the reparametrization identities to matrix models
in order to make the renormalization group approach more useful.
Work along this line is in progress.
%
\par
\vspace{3mm}
We thank B. Durhuus and T. Hara for an illuminating discussion and
M. Hirsch for a careful reading of the manuscript.
This work is supported in part by Grant-in-Aid for Scientific
Research (S.H.) and Grant-in-Aid for Scientific
Research for Priority Areas (No. 04245211) (N.S.) {}from
the Ministry of Education, Science and Culture.
\vspace{3mm}

\begin{thebibliography}{10}
%
\bibitem{BIPZ} E. Br\'ezin, C. Itzykson, G. Parisi and J.B. Zuber,
           {\it Commun.~Math.~Phys.~}{\bf 59} (1978) 59;
           D. Bessis, C. Itzykson and J.B. Zuber,
           {\it Adv.~Appl.~Math.~}{\bf 1} (1980) 109;
           V.A. Kazakov, I.K. Kostov and A.A. Migdal,
           {\it Phys.~Lett.~}{\bf 157B} (1985) 295;
           F. David, {\it Nucl.~Phys.~}{\bf B257} (1985) 45;
           V.A. Kazakov and A.A. Migdal,
           {\it Nucl.~Phys.~}{\bf B311} (1988) 171.
\bibitem{GRMI} D.J. Gross and A.A. Migdal,
           {\it Phys.~Rev.~Lett.~}{\bf 64} (1990) 127; 717,
           {\it Nucl.\ Phys.\ }{\bf B340} (1990), 333;
           E. Br\'ezin and V.A. Kazakov,
           {\it Phys.~Lett.~}{\bf 236B} (1990) 144;
           M. Douglas and S. Shenker,
           {\it Nucl.~Phys.~}{\bf B335} (1990) 635;
           D.J. Gross and N. Miljkovi\'c,
           {\it Phys.\ Lett.\ }{\bf 238B} (1990) 217;
           E. Br\'ezin, V.A. Kazakov and A. Zamolodchikov,
           {\it Nucl.\ Phys.\ }{\bf B338} (1990) 673;
           P. Ginsparg and J. Zinn-Justin,
           {\it Phys.\ Lett.\ }{\bf 240B} (1990) 333;
           G. Parisi, {\it Phys.\ Lett.\ }{\bf 238B} (1990) 209, 213.
%
\bibitem{DDK} F. David,
        {\it Mod.\ Phys.\ Lett.\ }{\bf A3} (1988) 1651;
        J. Distler and H. Kawai,
        {\it Nucl.\ Phys.\ }{\bf B321} (1989) 509;
        J. Distler, Z. Hlousek and H. Kawai,
        {\it Int.\ J. of Mod.\ Phys.\ }{\bf A5} (1990) 391; 1093.
\bibitem{BEKL} M. Bershadsky and I.R. Klebanov,
           {\it Phys.\ Rev.\ Lett.\ }{\bf 65} (1990) 3088;
            N. Sakai and Y. Tanii,
           {\it Int.\ J. of Mod.\ Phys.\ }{\bf A6} (1991) 2743.
%
\bibitem{DDSW} S. R. Das, A. Dhar, A. M. Sengupta and S. R. Wadia,
        {\it Mod.\ Phys.\ Lett.\ }{\bf A5} (1990) 1041;
        L. Alvarez-Gaum\'{e}, J.L.F. Barb\'{o}n and \v{C}. Crnkovi\'{c},
         preprint CERN-TH-6600-92 (July 1992).
\bibitem{BRHI} E. Br\'ezin and S. Hikami,
           {\it Phys.\ Lett.\ }{\bf 283B} (1992) 203;
           Ecole Normale preprint LPTENS-92-31 (September 1992).
%
\bibitem{BRZJ} E. Br\'ezin and J. Zinn-Justin,
        {\it Phys.\ Lett.\ }{\bf B288} (1992) 54.
\bibitem{WIKO} K.G. Wilson and J. Kogut,
        {\it Phys.\ Rep.\ }{\bf 12C} (1974) 75.
\bibitem{ALDA} J. Alfaro and P. Damgaard,
        {\it Phys.\ Lett.\ }{\bf B289} (1992) 342;
        V. Periwal,
        {\it Phys.\ Lett.\ }{\bf B294} (1992) 49;
        H. Gao,         Trieste preprint IC 302-92 (1992).
\bibitem{CARL} J. Carlson,
        {\it Nucl.\ Phys.\ }{\bf B248} (1984) 536;
        R.~Brustein and S.~P.~De~Alwis,
        University of Colorado preprint COLO-HEP-253 (1991).
%
\bibitem{NIYO} S. Nishigaki and T. Yoneya,
        {\it Nucl.\ Phys.\ }{\bf B348} (1991) 787;
        {\it Phys.\ Lett.\ }{\bf B268} (1991) 35.
\bibitem{DVKO} P. Di Vecchia, M. Kato and N. Ohta,
        {\it Nucl.\ Phys.\ }{\bf B357} (1991) 495;
        {\it Int.\ J. \ Mod.\ Phys.\ }{\bf 7A} (1992) 1391;
        P. Di Vecchia and M. Moshe,
        {\it Phys.\ Lett.\ }{\bf B300} (1992) 49.
\bibitem{AMP} A. Anderson, R. Myers and V. Periwal,
        {\it Phys.\ Lett.\ }{\bf B254} (1991) 89;
        {\it Nucl.\ Phys.\ }{\bf B360} (1992) 463.
\bibitem{ZJ} J. Zinn-Justin,
        {\it Phys.\ Lett.\ }{\bf B257} (1991) 335;
        Saclay preprint SPhT/91-054, 91-185 (1991).
\bibitem{FKN} M.~Fukuma, H.~Kawai and  R.~Nakayama,
        {\it Int.~J.~Mod.~Phys.~}{\bf 6A} (1991) 1385;
        R.~Dijkgraaf, E.~Verlinde and H.~Verlinde,
       {\it Nucl.~Phys.~}{\bf B348} (1991) 435.
\bibitem{ITOYAMAMATSUO}
         H.~Itoyama and Y.~Matsuo,
         {\it Phys.\ Lett.\ }{\bf B262} (1991) 233;
         L. Alvarez-Gaum\'{e}, C.~Gomez and J.~Lacki,
         {\it Phys.\ Lett.\ }{\bf B253} (1991) 56;
         A.~Mironov and A.~Morozov,
         {\it Phys.\ Lett.\ }{\bf B252} (1990) 47.
%
\end{thebibliography}
\end{document}